# Optimal Control-Based Eco-Ramp Merging System for Connected and Automated Vehicles

Zhouqiao Zhao, *Student Member*, *IEEE*, Guoyuan Wu, *Senior Member*, *IEEE*, Ziran Wang, *Member*, *IEEE*, and Matthew J. Barth, *Fellow*, *IEEE*

*Abstract*—Our current transportation system suffers from a number of problems in terms of safety, mobility, and environmental sustainability. The emergence of innovative intelligent transportation systems (ITS) technologies, and in particular connected and automated vehicles (CAVs), provides many opportunities to address the aforementioned issues. In this paper, we propose a hierarchical ramp merging system that not only generates microscopic cooperative maneuvers for CAVs on the ramp to merge into the mainline traffic flow, but also provides controllability of the ramp inflow rate, thereby enabling macroscopic traffic flow control. A centralized optimal control-based approach is proposed to smooth the merging flow, improve the system-wide mobility, and decrease the overall fuel consumption of the network. Linear quadratic trackers in both finite horizon and receding horizon forms are developed to solve the optimization problem in terms of path planning and sequence determination, where a microscopic vehicle fuel consumption model is applied. Extensive traffic simulation runs have been conducted using PTV VISSIM to evaluate the impact of the proposed system on a segment of SR-91 E in Corona, California. The results confirm that under the regulated inflow rate, the proposed system can avoid potential traffic congestion and improve mobility (e.g., VMT/VHT) up to 147%, with a 47% fuel savings compared to the conventional ramp metering and the ramp without any control approach.

## I. INTRODUCTION

### A. Motivation

Connected and Automated Vehicle (CAV) technology has been widely developed during the past decade. With on-board sensors such as cameras, radar, and LiDAR, CAVs can sense the surrounding environment and be driven autonomously and safely by themselves without colliding into other objects on the road. In addition, CAVs are able to communicate with each other or roadside infrastructure with vehicle-to-vehicle (V2V) communications or vehicle-to-infrastructure (V2I) communications, respectively, sharing information of vehicle status and infrastructure status (e.g., signal phase and timing). This enables CAVs to perform maneuvers in an efficient and collaborative manner.

As a fundamental scenario, merging (particularly at freeway on-ramps) attracts wide attention from many researchers due to the concerns of safety and mobility in the merging area, especially when the merging lane is relatively short and the merging vehicle cannot accelerate fast enough to reach a reasonable speed for merging into the main traffic flow. Also, since vehicles on the mainline often need to adjust their speed upon observing the merging vehicles during a

Z. Zhao, G. Wu, Z. Wang, and M. J. Barth are with Bourns College of Engineering, Center for Environmental Research and Technology (CE-CERT), University of California, Riverside, CA, 92507, USA (e-mail: zzhao084@ucr.edu; gywu@cert.ucr.edu; zwang050@ucr.edu; barth@ece.ucr.edu).

relatively short period of time, the intensive speed fluctuations and weaving maneuvers often lead to traffic congestion or shockwaves in the upstream segment, resulting in an increase of energy consumption of upstream vehicles. Moreover, uncontrolled inflow traffic from ramps to the highway may cause oversaturation conditions on the network and further aggravate the traffic congestion.

### B. Literature Review

Ramp metering is a widely used ramp merging management method, which utilizes traffic signals installed on highway on-ramps to regulate the inflow rate of traffic entering the mainline in response to prevailing mainline traffic conditions. Ramp metering usually consists of a two-phase signal light (red and green) together with a signal controller. It has proven to be a cost-effective operational strategy to improve mobility metrics and environmental impacts. Existing research has mainly fallen into three categories, namely rule-based approaches [1]–[6], control-based approaches [7]–[10], and learning-based approaches [11]–[15]. However, since ramp meters introduces stop-and-go driving for the ramp vehicles, it increases travel time and energy consumption. Also, ramp metering compresses the distances that merging vehicles have to adjust their speeds to merge into the mainline stream, potentially increasing the safety risk and fuel consumption.

In contrast to ramp metering, different kinds of connected vehicle technology and algorithms have been proposed and developed to address the ramp merging problems, through the introduction of coordinated motion control algorithms. Rios-Torres et al. [16], Scarinci et al. [17], and Zhao et al. [18] provided comprehensive reviews of the previous works regarding CAV-based cooperative ramp merging control. Milanes et al. developed a fuzzy-logic control method for vehicles to merge into the congested mainline, allowing flow speed changes of vehicles both on the ramp and on the mainline [19]. Marinescu et al. designed a slot-based approach for intelligent vehicles to merge from ramp to mainline, showing results of higher traffic throughput and decreased delay, compared to the baseline human-driven scenario [20]. The "virtual vehicle" methodology originated from Uno et al. [21], and has been further developed over the years by other researchers [22]–[24]. In this approach, CAVs on the mainline are projected on the ramp as virtual vehicles, where their information (distance to the merging zone, speed, acceleration, etc.) can be estimated and transmitted through V2V communications and/or V2I communications. Linear feedback controllers have been proposed for the ego vehicle in their studies to track the longitudinal movement of the virtual vehicle. Other than the aforementioned approaches for

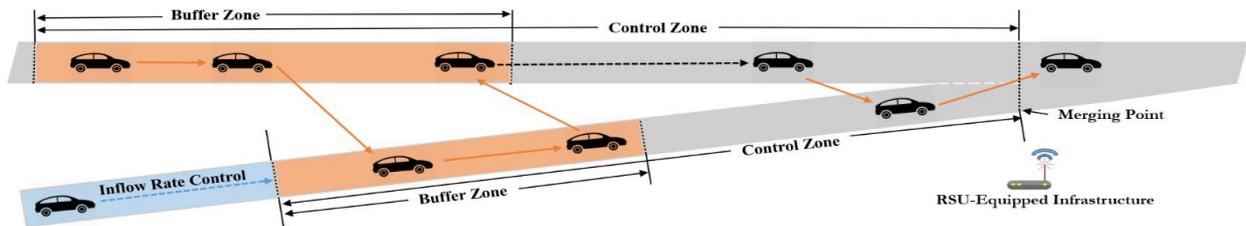
Fig. 1 Illustration of different zones of the proposed optimal control-based ramp merging system

ramp merging systems, optimal control has also been widely studied and implemented in this field of research. Rios-Torres et al. enabled the online coordination of merging vehicles by proposing an optimization framework together with an analytical closed-form solution [25]. Awal et al. developed a proactive optimal merging strategy to compute the optimal merging sequence of vehicles coming from both the mainline and the ramp [26]. Once the merging sequences of vehicles were defined, Ravari et al. presented a methodology to optimize the time-to-conflict-zone for vehicles to reduce their travel time [27]. Cao *et al.* proposed a model predictive control (MPC)-based path generation algorithm, which can generate the merging path for vehicles with real-time optimization [28]. Numerical simulation has been conducted based on traffic data recorded from a helicopter, and the results showed their proposed method can generate a cooperative merging path as long as the initial conditions of vehicles were reasonable.

*C. Contributions of this Paper*

Utilizing the traditional ramp metering methods, energy efficiency, safety, and mobility are the major problems that could be further improved. The contributions of our proposed system are related to the following issues of existing work in this research field:

- Energy Consumption: Utilizing CAV techniques, many researchers have been trying to design a cooperative control system between vehicles in the ramp merging area to improve mobility metrics. However, few of the above approaches take energy consumption as a major focus in the problem formulation.
- Entrance Sequencing: Although previous research proposed many sophisticated control methods, they failed to fully consider the entrance sequence of both mainline and ramp vehicles into the merging zone – the first-come-first-serve strategy and simplified estimated time of arrival (ETA) scheme are commonly used [25].
- Inflow Regulation: The cooperation of vehicles in the merging area proposed by existing research can only improve the local performance of the system. The unregulated inflow rate of the ramp vehicles can still lead to potential oversaturation of the highway network, thus increasing the risk of upstream congestion and traffic accident around the merging area.
- Microscopic Traffic Simulation: Existing research has not fully utilized comprehensive traffic simulation tools. Obtaining results from a snapshot of vehicles with numerical simulation cannot well represent the long term impacts on the traffic across a wide variety of scenarios.

This study aims to address the aforementioned problems by combining the ideas of both traditional ramp metering and CAV-based ramp control.

Our previous paper proposed the concept of a hierarchical system for corridor-wide ramp control [18]. At the corridor level of the system, a cooperative protocol can be introduced to calculate the system-wide optimal inflow rate for each on-ramp, given the estimate of macroscopic traffic conditions. The ramp level controller coordinates the maneuvers of CAVs locally at each ramp area and regulates the inflow rate accordingly. In this paper, we focus on the development of the ramp merging control subsystem (i.e., the ramp level of the proposed control system) by assuming a suggested ramp inflow rate to be given (from the upper level of the hierarchical system) as the input.

The structure of this paper is as follows. In Section II, we formulate the problem of highway ramp merging and discuss the hierarchical architecture of the developed system. Section III explains the details of the proposed methodology. The traffic simulation and associated results are presented in Section IV. Section V concludes the paper and discusses possible directions for future work.

II. Problem Formulation and System Architecture

*A. Problem Formulation*

Fig. 1. illustrates a typical highway-ramp merging area. However, different from the traditional ramp metering scenario which has a traffic light located at the end of the ramp, we consider the scenario with the following assumptions and specifications in this paper:

- All vehicles are CAVs whose information (e.g., position, speed, and acceleration) are precise and are shared via V2I or V2V communications. Further, their speeds can be fully controlled/executed by the acceleration/deceleration signals sent from a centralized processor.
- There is no communication delay or package loss in either V2V or V2I communications.
- Once the affected mainline vehicles are selected by the merging algorithm, they will not change lanes to preserve the number of controlled vehicles, or overtake other mainline vehicles to disturb the entrance sequence into the merging area.
- With the appropriate control, it is expected that the involved CAVs may avoid unnecessary stops (as mandatory by the ramp metering) before the completion of merging maneuvers, while the inflow rate on ramp or even the time headway between on-ramp vehicles can be well regulated.

Since we only control the longitudinal dynamics of the vehicles, we set up a one-dimensional coordinate system and map the positions of vehicles on both mainline and ramp to the system.

The dynamics of $n$ vehicles in the proposed ramp merging system can be given by:

$$\dot{p}_i = v_i, \dot{v}_i = u_i \quad (1)$$

where $i(\in [1,2,\ldots,n])$ is the vehicle index; $p$ and $v$ represent the position and speed of the vehicle, respectively; and $u$ denotes the acceleration of the vehicle, which acts as the input of the proposed system. If we define the overall system state as:

$$x = \begin{pmatrix} p_1 \\ p_2 \\ \ldots \\ v_1 \\ v_2 \\ \ldots \\ v_n \end{pmatrix}, \text{ and the observation as } y = \begin{pmatrix} p_1 - p_2 \\ p_2 - p_3 \\ \ldots \\ p_{n-1} - p_n \\ v_1 \\ v_2 \\ \ldots \\ v_n \end{pmatrix}$$

the system can be written as the following linear form:

$$\dot{x} = Ax + Bu$$

$$y = Cx \quad (2)$$

where $A$ is a $2n \times 2n$ system matrix of constant coefficients that describe the state transfer; $B$ is a $2n \times n$ control matrix of the coefficients that weigh the inputs; and $C$ is a $(2n-1) \times 2n$ output matrix.

Then, we formulate the optimization problem in the following quadratic form. The cost function is defined as the sum of the deviations of the measurements and control effort.

$$\min \quad J = \frac{1}{2} \sum_{k=0}^{N-1} \{(y_k - r_k)^T Q (y_k - r_k) + u_k^T R u_k\}$$

$$+ \frac{1}{2}(y_N - r_N)^T Q (y_N - r_N)$$

$$s.t. \quad x_{k+1} = Ax_k + Bu_k, y_k = Cx_k \quad (3)$$

$$Acc_{\min} \leq u_k \leq Acc_{\max}$$

$$((p_i)_k - (p_{i+1})_k) \geq Gap_{\min}$$

where $k$ is the time index of the finite horizon from 0 to N; $r_k$ is the gap and speed reference to be tracked; $Q$ and $R$ matrices define the weights of the objective function to be tuned, respectively, for the system outputs and inputs. $[Acc_{\min}, Acc_{\max}]$ is a feasible input (i.e., acceleration) range that the vehicles can achieve. $Gap_{\min}$ is the hard safety constraint to avoid the collision. If vehicle $i$ and vehicle $i+1$ are on the same lane (i.e., either both on the ramp or both on the mainline), this constraint should be held strictly. If vehicle $i$ and vehicle $i+1$ are on different lanes (e.g., one on the mainline while the other on the ramp), this constraint need to be held when they arrive at (or very close to) the merging area. It is noted that herein we simplify the safety constraint by borrowing the concept for legacy vehicles. CAVs may require more complex collision-free constraints (e.g., string stability issue for a platoon of CAVs) or even shorter $Gap_{\min}$ than legacy vehicles (from an individual CAV perspective) due to their quick reaction times.

*B. Control Zone and Buffer Zone*

To solve the problem, we first specify the roadway segment with two types of zones: control zone and buffer zone for the on-ramp and mainline, respectively, as shown in Fig. 1. In the control zones of both mainline and on-ramp, a centralized processor is employed to receive and process the incoming information from CAVs and send the control signals back to CAVs to achieve system-wide energy efficiency. Buffer zones (in orange) are located on the upstream portion of the control zones and are designed to continuously monitor the incoming vehicles and collect information to support subsequent control decisions.

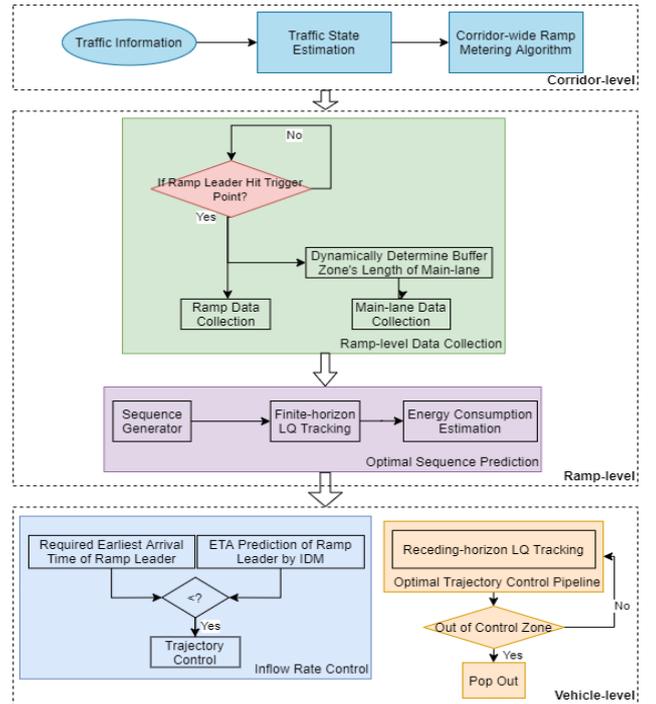

Fig. 2 Flow chart of the proposed ramp merging system

*C. System Architecture*

Fig. 2 illustrates the system architecture of the proposed highway on-ramp merging system. As described earlier, in this paper we focus on the development of the middle-level and bottom-level in the hierarchical system: ramp-level and vehicle-level, assuming the external input – the suggested ramp inflow rate is provided from the top level (or corridor-level).

*1) Ramp-level: data collection and vehicle merging sequence optimization.* In the data collection module, vehicles' information (such as position, speed, and lane index) is collected once they enter the buffer zones. The ramp buffer zone is predefined in this study based on the road conditions, while the length of the mainline buffer zone may vary with prevailing traffic conditions to guarantee proper number of mainline vehicles are controlled.

Because the controller is designed to regulate the vehicles to form a compact string, it is important to know the entrance sequence of the vehicles into the merging zone. Intuitively, the sequence would have impacts on the string-wise energy consumption. Instead of using a heuristic sequencing protocol, such as first-come-first-serve, we developed an optimal sequence determination module to identify the most energy efficient sequence. The module can exam all possible sequences, applied the finite-horizon linear quadratic (LQ) tracker to predicting the controlled speed profile of each involved vehicle, estimate the associated energy consumption, and select the least energy consumption sequence for the vehicle level control.

*2) Vehicle level: motion control of individual vehicles.* For the involved vehicles, the longitudinal motion controller is designed to be a receding horizon LQ tracker. To match the predicted energy consumption in the optimal sequence determination step, we use the *same controller* parameters as in the associated finite-horizon LQ tracker. Details of controller design will be presented in the following section.

### III. METHODOLOGY

In this section, we describe the detailed methodology for each key step in the flow chart of the proposed ramp merging system (shown in Fig. 2), including ramp-level data collection, optimal sequence determination, and vehicle motion control. We first define the concept of ramp leader, which is the key vehicle that triggers each event-based decision cycle and realizes the flow regulation. Ramp leader is the first vehicle that does not pass through the buffer zone and has not been controlled. More specifically, if no previous ramp vehicle is controlled by previous decision cycle, then the leading vehicle on ramp is the ramp leader, otherwise the first ramp vehicle that follows the last controlled vehicle is the ramp leader. Once the ramp leader passes the downstream boundary of the on-ramp buffer zone, a new ramp leader should be found upstream. Fig. 3 shows an example of the trajectories of vehicles in the system to illustrate how vehicles are grouped and controlled. The solid blue curves present the controlled trajectories of the ramp leader with the compliance of the inflow rates while the blue dash curves depict the predicted trajectories of the ramp leader under an Intelligent Driver Model (IDM, see [30]) approach; The yellow curves represent the trajectories of uncontrolled vehicles following the ramp leader; The orange curves show the vehicles under optimal control.

*A. Ramp-level Data Collection*

The buffer zone is designed to differentiate the involved vehicles within each control decision cycle for online implementation. In this study, the length of the on-ramp buffer zone is predefined, while the length of the mainline buffer zone is determined with the following equation:

$$L_{main} = \frac{\frac{q_{main}}{q_{suggested}} n}{d_{main}} \quad (4)$$

where $q_{main}$ is the mainline traffic flow known from corridor traffic condition; $q_{suggested}$ is the suggested on-ramp inflow rate assumed to be known; n is the number of on-ramp vehicles currently in the buffer zone; $d_{main}$ is the mainline density.

The vehicles in the buffer zones of both on-ramp and mainline would be controlled as a whole set until their travel through the merging area, which is shown as the orange curves in Fig. 3. Until the next ramp leader (in solid blue curves) reaches the downstream boundary of the on-ramp buffer zone, a new control decision cycle is initiated and a new set of involved vehicles is determined. It should be pointed out that the new control decision cycle does not mean the end of the previous control cycle. The control of the previous set of vehicles ends only if they all travel through the merging area. Depending on the prevailing traffic conditions, the number of involved vehicles in each set (during each control decision cycle) may vary and the control processes or multiple LQ tracker controllers for different vehicle sets may perform in parallel.

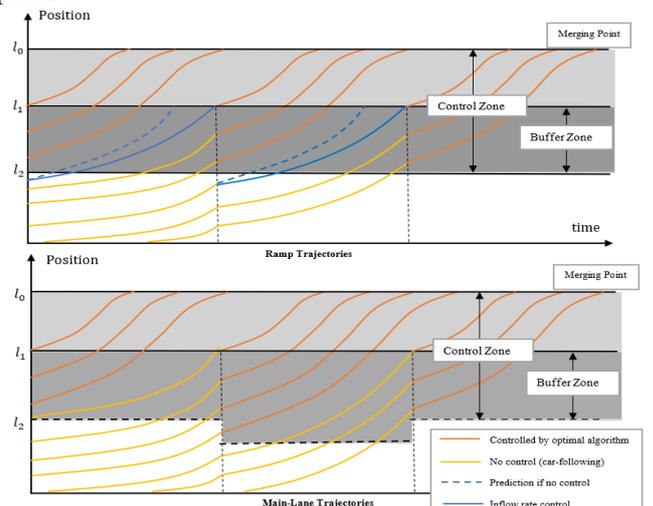

Fig. 3 Illustration of the vehicle's trajectories on ramp and the mainline

*B. Optimal Sequence Determination*

There are three sub-steps in the *Optimal Sequence Determination* process, including possible sequence generation, linear quadratic tracking, and energy consumption estimation. In this process, all the possible orders of the involved vehicles will first be generated. For each order, the optimal system inputs (acceleration of each involved vehicle) can be solved by an LQ tracker. Then based on the system dynamics, the speed profile can be calculated, and the energy consumption can be estimated by a microscopic vehicle fuel consumption model [29]. Each possible order is associated with one aggregated fuel consumption value (for all the involved vehicles), and the sequence with the least aggregated fuel consumption is picked as the optimal scenario. Vehicle-level will then use this order to control the vehicles motion.

*1) Possible sequence generation:* Given the assumption that all the involved vehicles within each control decision cycle would not change their lanes during the merging process, vehicles on the same lane can not overtake the preceding vehicles. Therefore, if there are $M$ mainline vehicles and $N$ on-ramp vehicles, the number of the possible sequence after merging equals to $P(M + N, N)$, where $P(\cdot)$ is the permutation operation. Because a large number of involved vehicles may lead to combinatorial explosion and make the system intractable for real-time implementation, the predefined length of the on-ramp buffer zone plays a key role to confine the control group to a reasonable scale. This can be considered as one way to balance the computational load.

*2) Linear quadratic tracking:* Based on the initial states, the finite-horizon linear quadratic tracking algorithm is able to generate the optimal solution in the designated finite time. The weight $Q$ and $R$ matrices are fine-tuned to keep the balance of tracking error and control input and also to hold the hard constraints. For better performance, the weighting factors for on-ramp vehicles and for those mainline vehicles are tuned independently. The solution is calculated iteratively as follows:

$$\begin{cases} S_N = C^T Q_N C \\ V_N = C^T Q_N r_N \end{cases} \quad (5)$$

$$\begin{cases} S_i = C^T Q C + A^T S_{i+1} - S_{i+1} B (R + B^T S_{i+1} B)^{-1} B^T S_{i+1} A \\ V_i = \{A^T - A^T S_{i+1} B (R + B^T S_{i+1} B)^{-1} B^T\} V_{i+1} + C^T Q r_i \end{cases} \quad (6)$$

$$\begin{cases} K_i = (B^T S_{i+1} B + R)^{-1} B^T S_{i+1} A \\ K_i^v = (B^T S_{i+1} B + R)^{-1} B^T \end{cases} \quad (7)$$

where Equation (6) is the discrete-time algebraic Riccati equation; $N$ is the predefined finite horizon; $i$ is the time index for each iteration; $K_i$ is the feedback gain and $K_i^v$ is the feed-forward gain. $S_i, V_i, K_i$, and $K_i^v$ are found iteratively backward in time. Then the solution is given by $\mu_i = -K_i x_k + K_i^v V_i$. With the control input $\mu_i$ and the system dynamic Equation (2), trajectories of all the vehicles can be calculated.

*3) Energy consumption estimation:* Based on the vehicle fuel consumption estimation model [29], the fuel consumption rate can be determined by the nonlinear function of current speed and acceleration:

$$f_V = b_0 + b_1 v + b_2 v^2 + b_3 v^3 + a(c_0 + c_1 v + c_2 v^2) \quad (8)$$

where $b_i$ and $c_i$ are the model parameters calibrated by different driving conditions; $v$ and $a$ are the speed and acceleration of the vehicles.

*C. Motion Control*

This module uses the results from the previous step to control the motion of the involved vehicles including vehicle cooperation control and ramp inflow rate control.

The vehicle cooperation controller chosen in this study is a receding-horizon LQ tracker for potentially online implementation. The advantage of receding-horizon LQ tracker than the previous finite-horizon LQ tracker is that it enables the closed-loop mechanism in the control system. At each rolling time window, the controller can update the initial states with the current state, and we only use the converged feedback gain and feed-forward gain to control the system. The Q and R parameters for this receding-horizon controller are selected to be the same as the ones used in the prediction step to get consistent results. When the constraints do not hold in a certain time step, the optimal solution will be recalculated by enlarging the current time window until the constraints are satisfied.

As described earlier, given the suggested ramp inflow rate, not all the ramp leaders should be controlled to enter the merging zone. The proper time of a ramp leader to reach th trigger point can be estimated as follow:

$$t_{proper} = \frac{n_{ramp}}{q_{suggested}} \quad (9)$$

where $n_{ramp}$ is the number of ramp vehicles selected for the previous decision cycle. Under the selected car-following model, if the ramp leader arrives at the trigger point earlier than this time, the ramp inflow rate would be higher than suggested. Here, we further assume the following behavior of the leader (to its predecessor if any) can be modeled by IDM [30], based on which we predict its ETA. The vehicle governed by the IDM presents a second-order dynamic shown as follow:

$$\dot{v} = a\left(1 - \left(\frac{v}{v_0}\right)^\delta - \left(\frac{s^*(v - \Delta v)}{s}\right)^2\right) \quad (9)$$

$$s^*(v - \Delta v) = s_0 + vT + \frac{v\Delta v}{2\sqrt{ab}} \quad (10)$$

where $v_0$ is desired speed; $s_0$ is minimum spacing; $T$ is the desired time headway; $a$ is the maximum vehicle acceleration; $b$ is the comfortable braking deceleration. The leader is controlled by a linear feedback controller if ETA is smaller than the suggested time until it hits the trigger point.

IV. SIMULATION STUDY AND RESULTS

In this section, we conduct a simulation study for the proposed ramp merging system with the microscopic traffic simulator PTV VISSIM [31] to validate the effectiveness of the system. Different from the numerical simulation that most of the previous research utilized, traffic simulation can offer more realistic real-time interaction between the equipped vehicles and other traffic in the network. This enables a better observation of the impact of the proposed system on the whole traffic over time. Through the *DriverModel* API, the behavior of the CAVs in the network can be controlled with the proposed algorithms. Uncontrolled vehicles in the network are modeled by the default behavior in VISSIM. The simulation network (see Fig. 4) is built based on the California State Route 91 (SR-91) East, with a focus on the Serfas Club Dr on-ramp in Corona California[1].

In our simulation, the conventional ramp metering system and the ramp without any control approach are introduced for comparison. Only longitudinal control is considered in the simulation, while the default lane change model is used for lateral control. Based on the observation, the merging area capacity is around 1,800 passenger car unit/hour/lane (pcu/hr/ln). According to this, the ramp inflow rate is

---

[1] The illustration video can be found at
https://www.youtube.com/watch?v=pcmEtsyVH58.

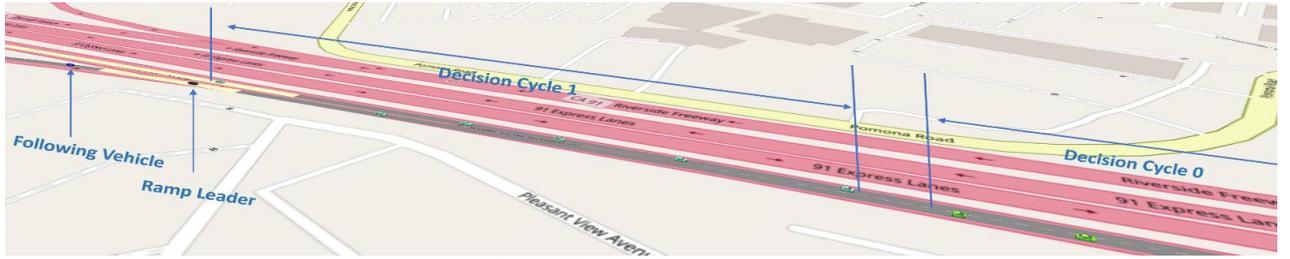
Fig. 4 PTV VISSIM simulation environment

dynamically adjusted to regulate the overall traffic flow not to exceed the capacity. For a fair comparison, the baseline ramp metering rate is also set to match the inflow rate that is regulated by the proposed system. The desired speed for mainline/merging traffic is 73.8 mph. The initial speed of on-ramp vehicles while entering the control zone is 33.5 mph; $Acc_{max}$ equals to $8.2 ft/s^2$; $Acc_{min}$ equals to $-9.8 ft/s^2$; and $Gap_{min}$ equals to 2s headway times the initial speed $v_{i0}$.

We consider two scenarios based on different traffic conditions. Each scenario contains two phases, lasting 600s respectively. For mainline traffic, 1,600 pcu/hr/ln is considered as heavy, and 1,200 pcu/hr/ln is as moderate. For ramp traffic, 500 pcu/hr/ln is considered as heavy, and 300 pcu/hr/ln is as moderate. Table I shows the settings of simulation scenarios in this study.

TABLE I. SCENARIO MATRIX

|  | Phase 1: 0-600s | | Phase 2: 600-1200s | |
| --- | --- | --- | --- | --- |
|  | Mainline Inflow | Ramp Inflow | Mainline Inflow | Ramp Inflow |
| Scenario 1 | 1600 | 500 | 1200 | 300 |
| Scenario 2 | 1600 | 300 | 1200 | 500 |

The simulation results measured by the mobility metric are shown in TABLE II and TABLE III. The mobility performance is measured by network efficiency,

$$Q = \frac{VMT}{VHT}$$

where *VMT* is the total vehicle-miles traveled in the network; and *VHT* is the total vehicle-hours traveled in the network accordingly.

In *Scenario 1*, the heavy traffic of both mainline and ramp in phase 1 rapidly caused the congestion in the network for both ramp metering case and no control case. At each time when the ramp vehicle merged, a shockwave was generated and spread to the upstream traffic, which eventually evolved to stop-and-go traffic along the mainline. In addition, the consistent shockwaves impeded the recovery of congestion, which led to low mobility of the network. As shown in TABLE II, the overall mobility of no control case has only 27.96 mph. Although ramp vehicles have relatively high mobility, their uncooperative behaviors severely influenced the mainline vehicles. On the other hand, in the ramp metering case, since the ramp inflow rate was regulated, a less significant impact was involved on the mainline. The mainline mobility in this case was 41.81 mph, much better than the no control case. However, the ramp metering operation severely limited the mobility on ramp, and the extremely high frequency of stop-and-go maneuvers at very low speed also caused significant energy waste for vehicles. As to the case of the proposed optimal control, the cooperation led to the highest overall mobility (including both mainline and ramp), which is improved by 109.6% and 147.5%, respectively, compared to the ramp metering case and no control case. In terms of fuel consumption, the proposed system outperforms both the ramp metering case and the no control by 47.5% and 45.5%, respectively.

In *Scenario 2*, since the heavy traffic of mainline and ramp were staggered, it was a generally moderate traffic condition compared to *Scenario 1*. Therefore, the mobility performance is better for all cases. The proposed optimal control system still achieved the best mobility, with improvement of Q by 64.1% compared to the ramp metering case and 141.8% compared to the no control case. In terms of fuel consumption, the proposed system also outperformed both the ramp metering case and the no control by 40.9% and 46.9%, respectively.

TABLE II. SIMULATION RESULT FOR SCENARIO 1

|  |  | Mobility (mph) | Fuel Consumption (mpg) |
| --- | --- | --- | --- |
| **Optimal Control** | Overall | 69.19 | 30.87 |
|  | Mainline | 72.70 | 34.52 |
|  | Ramp | 59.37 | 22.66 |
| **Ramp Metering** | Overall | 33.01 | 20.93 |
|  | Mainline | 41.81 | 23.94 |
|  | Ramp | 17.73 | 13.78 |
| **No Control** | Overall | 27.96 | 21.22 |
|  | Mainline | 25.14 | 21.73 |
|  | Ramp | 51.65 | 19.37 |

TABLE III. SIMULATION RESULT FOR SCENARIO 2

|  |  | Mobility (mph) | Fuel Consumption (mpg) |
| --- | --- | --- | --- |
| **Optimal Control** | Overall | 70.45 | 31.26 |
|  | Mainline | 73.64 | 34.40 |
|  | Ramp | 60.11 | 22.91 |
| **Ramp Metering** | Overall | 42.92 | 22.19 |
|  | Mainline | 54.96 | 25.44 |
|  | Ramp | 23.38 | 14.56 |
| **No Control** | Overall | 29.14 | 21.28 |
|  | Mainline | 26.32 | 21.74 |
|  | Ramp | 51.77 | 19.56 |

V. CONCLUSION AND FUTURE WORK

In this paper, we proposed a hierarchical ramp merging system for CAVs. The system not only coordinates the vehicles in the ramp merging area to achieve a safer, smoother, and more efficient traffic flow but also can regulate the ramp vehicles' inflow rate which has the potential to leverage the corridor-wise efficiency by integrating with effective perimeter control on multiple ramps. We developed a ramp-level data collection logic that can determine the right set of vehicles for online control and collect the associated information based on the prevailing traffic conditions. Unlike most existing studies using simple sequencing protocol (e.g., first-come-first-serve), we used a finite LQ tracker to identify the optimal merging sequence in terms of energy consumption. A receding horizon LQ tracker with the same parameters was used for optimal motion control. The traffic simulation results verified the effectiveness of the proposed system. Even in the scenario that traffic flow is more than the freeway capacity, the smoothed vehicle trajectories can lead to a placid speed reduction without shockwave or congestion. It should be pointed out that such multi-stage optimization framework tries to balance the problem simplification and system optimality. It cannot theoretically guarantee the global optimal solution to system-wide efficiency.

An ongoing research direction is to develop the upper-level corridor-wise perimeter control algorithm which can provide the optimal ramp inflow rate (cooperatively) for each involved individual ramp along the corridor. In addition, more practical considerations on mixed traffic scenario would be one of our future steps.

ACKNOWLEDGMENT

This study is funded by the National Center for Sustainable Transportation (NCST). The contents of this paper reflect only the views of the authors, who are responsible for the facts and the accuracy of the data presented herein.